\documentclass[12pt]{article}%
\usepackage[dvips]{graphicx}
\usepackage{amsmath, amsthm, amssymb}
\usepackage{amsfonts}
\usepackage{color}
\usepackage[bookmarks=false,colorlinks=true,urlcolor={blue},pdfstartview={XYZ null null 1.22}]{hyperref} 
\long\def\symbolfootnote[#1]#2{\begingroup%
\def\thefootnote{\fnsymbol{footnote}}\footnote[#1]{#2}\endgroup}
\begin{document}
\definecolor{high}{rgb}{0.3,0.3,0.9}
\newcommand{\mycolor}{high}
\title{Abridged Petri Nets}
 \author{Vitali  Volovoi}
  \maketitle
\begin{abstract}
A new graphical framework, Abridged Petri Nets (APNs) is introduced for bottom-up modeling  of complex stochastic systems.  APNs  are similar  to  Stochastic Petri Nets (SPNs) in as much as they both rely on component-based representation of system state space, in contrast to  Markov chains that explicitly model the states of an entire system. In both frameworks, so-called tokens (denoted as small circles) represent individual entities comprising the system; however,  SPN graphs contain two distinct types of nodes (called places and transitions) with transitions serving the purpose of routing tokens among places. As a result,  a pair of place nodes in SPNs can be linked to each other only via a transient stop, a transition node. In contrast, APN graphs link place nodes directly by arcs (transitions), similar to state space diagrams for Markov chains, and separate transition nodes are not needed. 
 Tokens in APN are distinct and have labels that can assume both discrete values (``colors") and continuous values (``ages"),  both of which can change during simulation. Component interactions are modeled in APNs using triggers, which are either inhibitors  or enablers (the inhibitors' opposites). Hierarchical construction of APNs  rely on using stacks (layers) of submodels with automatically matching color policies.  As a result, APNs provide at least the same modeling power as SPNs, but, as demonstrated by means of several examples, the resulting models are often more compact and transparent, therefore facilitating more efficient  performance evaluation of complex systems. 
\end{abstract}
\section{Background}
The paper introduces Abridged Petri Nets (APNs), a new graphical framework for modeling the stochastic behavior of complex systems that consist of multiple interacting components. This framework can be considered as a derivative of Stochastic Petri Nets (SPNs)~\cite{Marsan1990} that aims at retaining SPNs versatility in terms of modeling power, while streamlining the choice of the modeling building blocks. The visual clutter and often confusing choices that are often perceived as the major obstacle to the larger success of SPNs are reduced~\cite{Bowden2000}, resulting in simpler and more transparent models that can be built using only graphical interface.

The  main distinguishing feature of the proposed framework is not {\em the addition of} new objects, but on contrary {\em removing}  one of the most fundamental types of objects in SPNs---transitions---as distinct nodes  of graphs.  Transitions  as the hubs or junctures of tokens' movements (depicted as rectangles) provide an ingenuous  mechanism for modeling components' interactions in classical SPNs.
The original idea of Petri nets was conceived  in the context of chemical reactions~\cite{PetriReisig2008}, where the merging (joining) and splitting (forking) of entities is quite common, which might explain the fundamental role of transitions that provide a direct means of modeling these processes. The merging and splitting is also important in the context of informational flows. In other applications such events occur as well, but equally if not more important are the changes that occur to entities individually. In fact, there are clear competitive advantages to modular structuring of complex systems~\cite{Simon2002}, which might explain its prominence in both natural and engineering systems leading to so-called near-decomposibility~\cite{Courtois1977}. With this consideration in mind, entities that comprise complex systems can be considered as operating mainly  independently of each other, with interactions occurring relatively rarely. However, when these interactions do occur, they are critical to the system's behavior.

Therefore, it would be natural to use a modeling framework that reflects this near-decomposability and  considers independent (parallel) behavior as the default while explicitly focusing attention on the interactions. Indeed this is the main underlying principle of the described framework.  The joining and forking  of entities represents important mechanisms for modeling a system's interactions,  but, as described in the examples in this paper, quite often the actual mechanisms are related to  enabling or inhibiting state transitions, while joining and forking provide a means of modeling such mechanisms. The difference is subtle and often not noticeable when tokens are indistinguishable (as in classical SPNs), but the consequences become significant when the entities represented by tokens have distinct identities (labels, such as colors~\cite{Jensen1993}).
 When a doctor is available to see the patient, the two entities are not actually joined into a single patient-doctor entity, which is then split back into two separate entities when the visit is over. This is simply a convenient SPN representation, but it leads to complications when we want to distinguish individual patients and doctors. The label of the token representing patient-doctor would need to contain information permitting the proper way of splitting the token into two and restoring the identity (and attributes) of  each member of the pair and routing them into appropriate places. While this is possible, it inevitably leads to  labeling and routing rules that are quite complex, often hindering visual understanding of the process.
 
  In contrast, inhibitors and their opposites, enablers (collectively referred as triggers, {\em cf.} test arcs~\cite{Christensen1993}), are free of those complications. Inhibitors are a popular extension of SPNs as they provide a so-called ``zero-test",  thus elevating  the modeling power of the framework to that of a Touring machine. However, both inhibitors and enablers (non-zero test arcs) are  usually considered supplementary mechanisms for modeling interactions  rather than  a replacement for fork-joining capabilities provided by transitions. Historically, inhibitors were viewed with a certain degree of skepticism by the Petri net community, as they make traditional analysis of structural properties (such as reachability analysis) more complex. However, the latest view of this drawback of inhibitors is not as straightforward as the new algorithms  can successfully handle inhibitors~\cite{Ciardo2004}. In addition, there is a sufficient number of applications (e.g., modeling  failure and maintenance processes of complex systems) where the state-space of the problem is relatively well understood and the main utility of the modeling consists of quantitative performance evaluation of the system.
 This paper is the first (to the best of the author's knowledge) to rely on triggers  as a sole alternative mechanism  for describing components' interactions.  This enables direct connectivity of places, and when combined with  hierarchical representation,  the result is a compact yet powerful modeling framework.

 The paper is organized as follows: the rest of the section discusses Markov chains and Stochastic Petri Nets and introduces several representative example. Section 2 introduces APNs, with hierarchical issues discussed in Section 3, followed by conclusions.

\subsection{Markov chains}
State-space diagrams provide a convenient  graphical way of depicting the behavior  of nondeterministic systems. Markov chains are the simplest and most popular kind of state-space diagrams, with applications ranging from their original use by Andrey Markov (for modeling the relative frequencies of vowels and consonants in Alexander Pushkin's novel {\em Eugene Onegin}) to evaluating web page ranks (in the PageRank algorithm invented by Google's founders)~\cite{Hayes2013}. Despite their deserved popularity, Markov chains are prone to  ``state-space explosion"---they scale poorly as the number of components increases that comprise the system. 

Let us consider a simple example that illustrates this issue, with a household consisting of  two family members (later referred to as  customers), $m_1$ and  $m_2$,  and a car, $c_1$.  Each customer can be in one of three states: not needing a car ($N$), driving a car ($D$), or waiting for a car ($W$); we assume that being a passenger in the car qualifies as state $N$. 
The inputs to this model could include the usage pattern of each customer, {\em e.g.,} the frequency and  duration of trips, as well as the car properties, { \em e.g.,} the  frequency of breakage and the duration of the repairs.  The outputs would be the ``performance" measures or metrics of this ``system," such as the frequency  and duration of unsatisfied demand.  These measures can help to make educated decisions  about changes to the system,  {\em e.g.,}  whether it makes sense to  get a second car.

First, let us consider a situation where the car  can be in one of two  states: idle ($I$) or used ($U$). 
This ``system" consists of three entities (two customers and a car), so there are $3\times 3\times 2=18$ possible perm{utations of the components' states that can define the state the entire system.  However, not all of those permutations constitute  feasible system states, so a Markov model would have only five rather than  $18$ states. Figure~\ref{markov1} shows the corresponding Markov chain diagram. Here the state of each entity is represented by the corresponding capital letter (so, for example, the DNU state of the system implies that first customer is driving, the second is not needing the car, and the car is used). 
Introducing the possibility that the car can be broken (be in state $B$) extends the size of the diagram to nine states (as shown in Figure~\ref{Markov2}). For a scenario with $n$ customers and $k$ cars, the number of system states can be calculated as follows: at any given time we can have $0\leq m \leq n$ members of the family who need cars (demand) and  $0\leq l \leq k$  cars that are not broken (supply). For each $m\leq l$, the number of distinct states corresponds to the number of ways we can select $m$ driving members of the family: $N(m,l)=\binom {n}{m} \binom {l}{k}$. 
If $m> l$, then $m-l$ members of the family are  waiting for a car ({\em i.e.,} in state $W$), so the corresponding number of states will be $N(m,l)=\binom {n}{m} \binom {m}{l}  \binom {l}{k}$.  The total number of states will be given by the following formula:
\begin{equation}
\sum_{m=0}^{n}{\binom {n}{m} \left[\sum_{l=0}^{m-1}{\binom {m}{l}  \binom {l}{k}}+\sum_{l=m}^{k}{ \binom {l}{k}}\right]}
\label{MarkovStates}
\end{equation}
For  a family of four with two cars, the number of possible states is $111$, and for a family of five with three cars, that number is $589$.  If we consider a fleet of $10$ cars with $20$ customers, the number of states is $451,417,560,951$---or over $451$  billion.   
\begin{figure}[ht]
\begin{center} 
\includegraphics[width=3.0in]{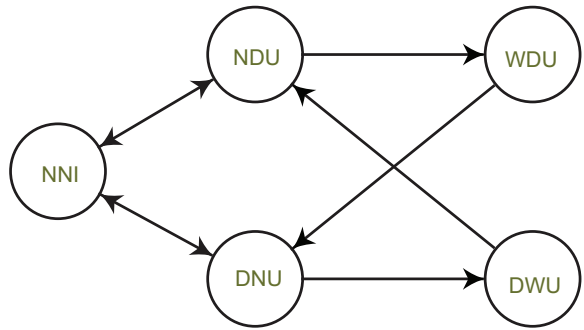}
 \end{center}
\caption{Markov chain diagram for two customers and one car}
\label{markov1}
\end{figure}

 \begin{figure}[ht]
\begin{center} 
\includegraphics[width=3.0in]{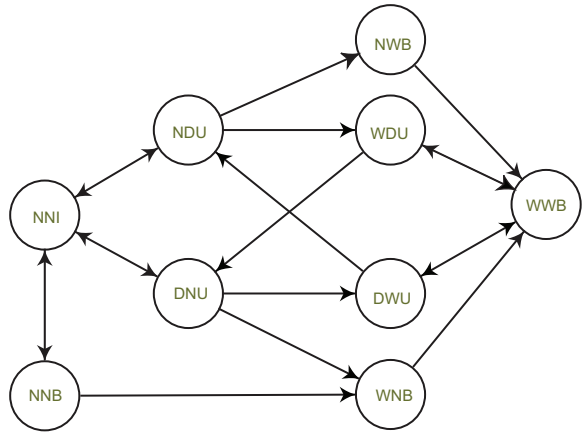}
 \end{center}
\caption{Markov chain diagram for two customers and one car, including the possibility that the car is broken}
\label{Markov2}
\end{figure}

This rapidly growing state-space size  can be avoided using symmetry considerations if the customers  or the cars are not distinguished among themselves. Indeed, for two customers and one car, the number of states reduces to three and six for Figures~\ref{markov1},~\ref{Markov2}, respectively, as states such as $DNU$ and $NDU$ can be merged together.  However, very often one does want to track individual performance (Dad might need the car more often for longer trips, and Car A is old and breaks more often). On the other hand, the state space will be even larger if we want to distinguish which car is driven by a given customer: if there are $q$ cars driven at any given time, there are potentially $q!$ possible combinations, although only some of those combinations might be feasible (for example, one can have a list of  preferred cars for each member of the family and track how their preferences are satisfied).
\subsection{Petri nets}
 The above example represents   a  general problem with customers generating service demand  and cars representing entities (servers)  that service (satisfy) the demand. The meaning of both customers and servers is domain-dependent. A partial, but not exhaustive, listing of those pairs includes:
\begin{itemize}
\item  Packets of information (customers) and computer servers or routers (servers) in computer applications
\item Tasks (customers) and resources (servers) in business processes
\item Callers (customers) and tellers (servers) at the call centers
\item Passengers or freight (customers) and transportation and storage resources (servers) in logistics
\end{itemize}
As a result,  matching demand and supply problems are at the core of modeling needs with Petri Nets~\cite{Marsan1990}. 

Noting that  alphabetical notations based on the states of the components were used for identifying a system state, it is natural to employ some sort of equivalent of an alphabet in a graphical description of the system.  Petri nets  effectively implement this idea by modeling the states of individual components rather than the explicit states of the entire system. In Petri nets, Markov 
chain-state diagrams are complemented by two new types of objects: first, small filled circles (called tokens) denoting individual components are introduced, each placed inside of one of the  larger hollow circles that denote the potential states of those components (the latter entities are named ``places" as opposed to ``states" in Markov diagrams). 
Second, in order to model interactions among components, the tokens are routed among places via intermediate stops or junctures, called transitions, which are denoted with solid rectangles. Any two places cannot be connected by an arc directly; instead they must be connected through a transition. The number of input and output arcs does not need to coincide, enabling the merging and splitting of token routes. The timing of state changes can be modeled by specifying time delays for transition ``firing": an action that removes tokens from all input places for the transition and deposits tokens into its output places. Such Petri nets are  timed Petri nets, or, more specifically, Stochastic Petri Nets (SPNs)~\cite{Balbo2007,Haas2002}, when  delays can be nondeterministic  and follow a specified distribution (no limitations on the associated  types of distributions are considered here). 

 Figure~\ref{spn1} depicts an SPN for two customers and one car. Using standard notations for SPNs, thin rectangles denote immediate transitions that incur no delays. 
We note that the SPN in Figure~\ref{spn1} consists of three groups corresponding to each component of the system, and if there are tokens in the places ``car needed" and ``car available," those two tokens are merged into a single token deposited into the place ``car used." This simple model reflects the fundamental feature of SPNs in modeling the coordinated behavior of system components: tokens that represent the car and the driver are merged into a single ``car-driver" token while driving takes place, and split into separate tokens again when the driving is complete.

The model has eight states (places), but clearly the complexity of the model is determined not only  by the number of places but also by the number of transitions (eight),  connecting arcs ($20$), and tokens ($3$). In fact, this model does not take into account the possibility that the car can break while driving (so each trip is completed); to include this possibility, we need to to include two extra transitions as shown in  Figure~\ref{spn2}. Now we have ten transitions and $26$ arcs (we note that the corresponding Markov model has $23$ arcs).

\begin{figure}[ht]
\begin{center} 
\includegraphics[width=3.0in]{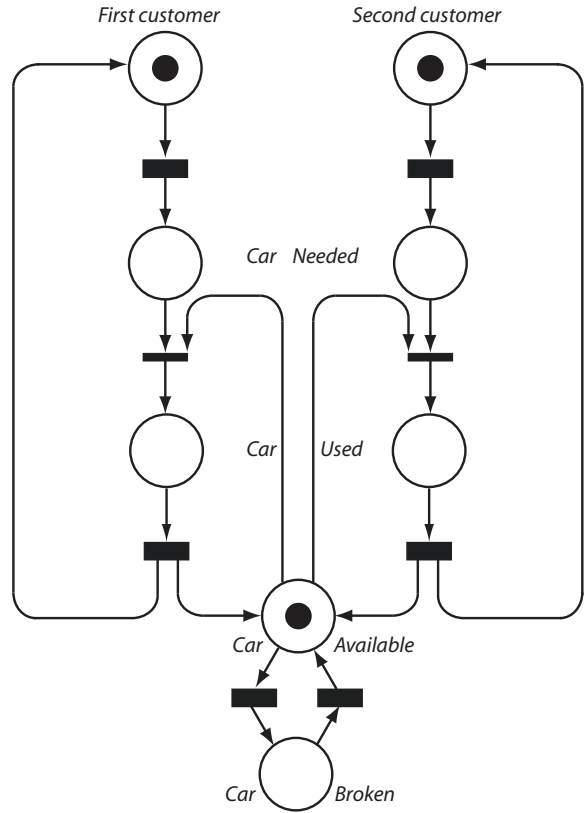}
 \end{center}
\caption{Stochastic Petri Net diagram for two customers and one car (car assumed not to break during  trips). }
\label{spn1}
\end{figure}

\begin{figure}[ht]
\begin{center} 
\includegraphics[width=3.0in]{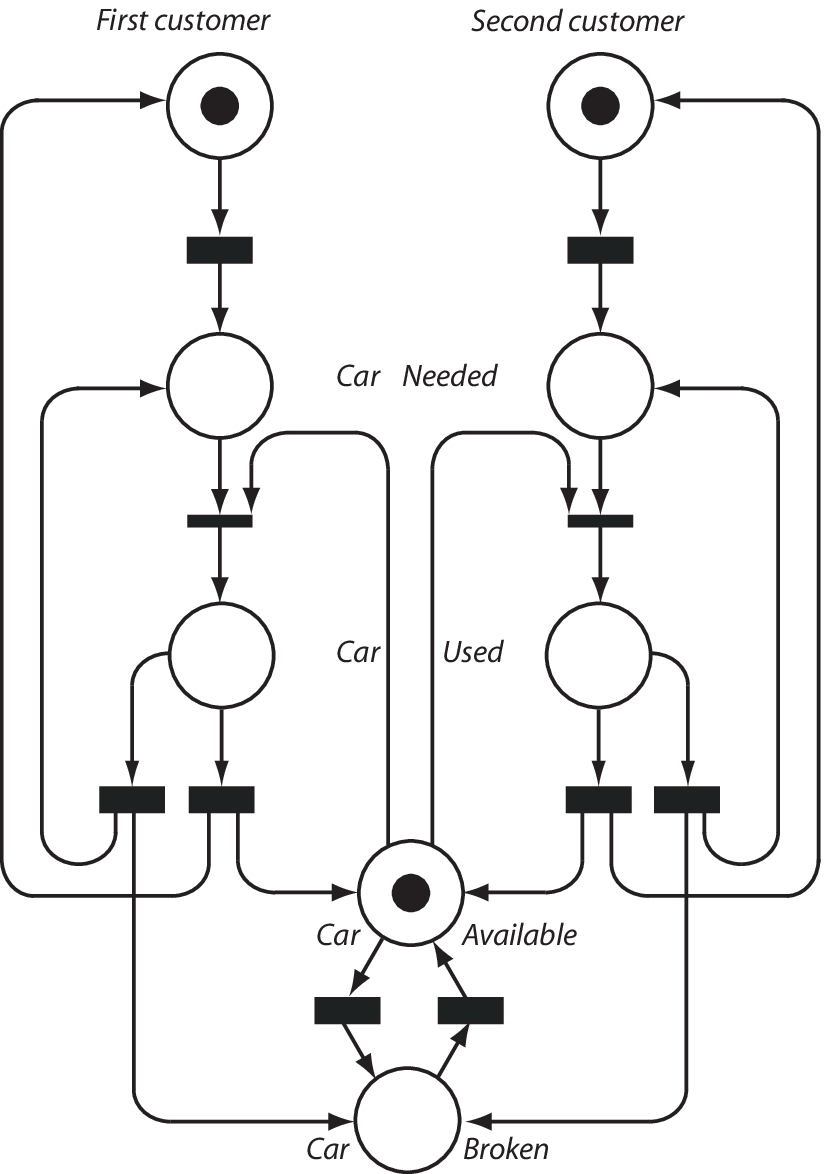}
 \end{center}
\caption{Stochastic Petri Net diagram for two customers and one car, including the additional possibility that car can also  break during the trip.}
\label{spn2}
\end{figure}

Such SPN  models scale better than Markov chains (which explains the fact that they were originally used as pre-processors for creating Markov chain models~\cite{Trivedi2002}). Indeed, there would be three places required for each customer and two places for each car, so the number of places in the model with $20$ customers and $10$ cars would be only $80$, which is certainly an improvement over $451$ billion states. However, the web of connecting arcs will be so convoluted that the resulting model is still  too complex to be of practical use for conveying the system behavior visually (referring back to Figure~\ref{spn2}, one can envision  $20$ segments  of the net similar to the two depicted at the top of the net, and $10$ segments similar to the one to the bottom, with each of the $10$ segments  at the bottom being connected to each of the $20$ segments at the top).

\subsection{Using high-level extensions of Petri nets}

Noting that the subnets for each customer are similar, it is tempting to use only one of the subnets and represent each customer by a different token within the same net.  At least two modifications are needed to enable such modeling: 
\begin{enumerate}
\item Parallel processing of tokens by transitions: we need to define the behavior of the net when there are multiple tokens in the same place. While some SPNs use  ``single-server" enabling (when tokens are enabled and fired one a time), here the multiple enabling (or ``infinite-server," to be precise~\cite{Balbo2007}) is preferable, so that each token's eligibility for moving to a new place is assessed in parallel ({\em e.g.,} two customers might want to drive a car simultaneously). While it is possible to incorporate both single and multiple servers within  the same framework, single servers can be easily represented using multiple servers, 
so only multiple servers are used herein.
\item  Colors: As discussed in the context of Markov chains, the state-space explosion is primarily caused by the need to account for differences in component behavior. If a token representing a component is traveling within a distinct subnet ({\em e.g.,}  Fig.\ref{spn2}), we can incorporate the differences by appropriately adjusting the properties of individual transitions for each subnet. Introducing colors to Petri nets~\cite{Jensen1993} allows for the transition properties to be color-dependent, and so components can maintain their differences while the corresponding tokens travel through the same subnet.
\end{enumerate}
 The resulting network is shown in Figure~\ref{spn3}. It  looks more compact and scalable. However, implementing such a model requires a fairly complex definition of what ``color" means. The original concept of a token's color~\cite{Jensen1993} allowed for complex attributes to be assigned to tokens, as eel as the associated means of transforming those attributes by means of complex ``inscriptions" (often elaborate formulae specified for transitions), which allowed for powerful modeling  at the (substantial) expense of reduced model readability and visual transparency.
 This can be contrasted with the simplest and most intuitively appealing implementation of a ``color" is an integer assigned to a token that can  be visualized by the corresponding color (as shown in  Figure~\ref{spn3}, tokens' integers can be  depicted explicitly as well, which is convenient for black-and-white implementations).  One can  observe that the merging and splitting of tokens at transitions causes some bookkeeping difficulties in terms of tracking individual tokens' identities (colors). For example, when the car is used, there is a token that represents the first customer using the car, and the following transition should ``know" the past of that merged token to restore the original  token that represents the customer and route it to the top place. If we have multiple cars and family drivers, the permutations will multiply---requiring a matrix of attributes, which would explicitly stipulate  complex rules governing the merging of colors and then splitting them back.
\begin{figure}[ht]
\begin{center} 
\includegraphics[width=2.0in]{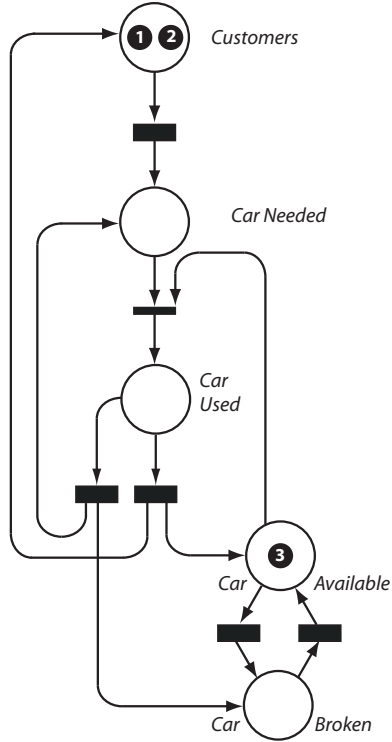}
 \end{center}
\caption{Colored Stochastic Petri Net diagram for two customers and one car. Integer labels (colors) are directly shown inside each token.}
\label{spn3}
\end{figure}
\section{Abridged Petri Nets (APN)}
These difficulties of tracking individual tokens through transitions can be avoided by allowing only a single input and a single output through a transition. As a result, the need for a separate node that constitutes junctures for routing nodes among places is eliminated, and places can be directly connected by arcs. Interaction among different entities in the system is exclusively modeled using triggers:  inhibitors and their opposites, enablers.

Inhibitors provide a ``zero test" capability and are known to increase the modeling power of Petri nets to that of a Touring machine~\cite{Balbo2007, Haas2002}. They are depicted as arcs originating at a place and terminating at a transition with a hollow circle.  An inhibitor of multiplicity $k$ disables a transition at which it terminates if the number of tokens in its input place is at least $k$.  An enabler  (depicted  as an arc originating at a place and terminating at a transition with a filled circle)  is defined in the opposite way:  a transition is disabled unless an enabler of multiplicity $k$ has at least $k$ tokens in its input place~\cite{Volovoi2006} (enablers are effectively test arcs ~\cite{Christensen1993}). Test arcs are used in system biology modeling~\cite{Matsuno2003}, where they are denoted with directed dashed arcs; the  current terminology  is chosen to emphasize  the fact that enablers are the opposite of inhibitors.
It can be shown that the combination of enablers and inhibitors with direct transitions between places allows the modeling of any system that can be modeled using Petri nets with multiple inputs and outputs to a transition (in other words, the modeling power is not reduced). 

In the past,  SPNs used transitions as hubs for routing tokens as the main mechanism for modeling component interdependence with  triggers (mainly inhibitors)  as a supplementary means for modeling the interdependence. In contrast,  the described framework  relies solely on the triggers, and as a result can  connect places directly, thus disposing of the hubs (transitions) altogether.

The result resembles a hybrid between traditional Markov chain-state charts and Petri nets: transitions connect places directly (similar to Markov chains), but tokens are present to represent individual components of the system (similar to Petri nets). Importantly, the tokens have discrete labels (colors) as well as continuous labels (age)~\cite{Volovoi2004}. As shown below, when combined with hierarchical construction capabilities, the  framework provides compact and visual representation of complex interactions in large-scale systems. We refer to the resulting framework as ``Abridged Petri Nets (APNs)".

\subsection{Properties of APNs}
\begin{enumerate}
\item An APN is defined  as a network of  places (denoted as hollow large circles) that are connected by directed arcs (transitions). Changes in the system's state are modeled by transition firing: {\em i.e.,} moving a token from  the transition's input place to its output place.  The combined position of tokens in the net at any given moment represents marking of the net and fully specifies the modeled system.
\item Each transition has no more than a  single input and single output place (a transition can also have no input place, providing a source of new tokens every time it fires, or it can have no output place, providing  a sink for tokens; upon the firing of such  a transition, a token is removed from the net).  
\item Each token can have a discrete label (color) that can change when token moves. In addition, tokens have continuous labels (ages) that can change both when tokens move, and with the progression of time while a token stays in the same place~\cite{Volovoi2009}. 
\item  A transition is enabled or disabled based on  the  combined marking of the input places of the associated triggers  (inhibitors and enablers). Inhibitors are depicted as arcs originating at a place and terminating at a transition with a hollow circle.  An inhibitor of multiplicity $k$ disables a transition that it terminates at if the number of tokens in its input place is at least $k$.  An enabler (depicted  as an arc originating at a place and terminating at a transition with a filled circle) is opposite to an inhibitor:  a transition is disabled unless an enabler of multiplicity $k$ has  at least $k$ tokens in its input place. Triggers can be color-specific (and therefore enable only tokens in a place of a certain color, or act only if there is a specified number of tokens of a given color in their respective input places, or both). 
\item  Transitions  have color- and age-dependent policies that specify the delay between the moment when the token is enabled and when it is fired.
If a token-transition pair is enabled, a firing delay is specified based on the combination of token and transition properties. If the token stays enabled throughout the delay, after this delay expires the token is fired. If there are multiple enabled tokens in the same place, they all can participate in the firing ``race" in parallel. Similarly,  the same token can be involved in a race with several transitions. If a token-transition pair is disabled, the firing is preempted  (however, the aging label of the token can change as a result of being enabled for a finite amount of time). 
\item The delays can be deterministic (including zero delay) or follow  any specified  random distributions. 
 \item The performance of the system is based on the statistical properties of marking of the system, and can be evaluated using discrete-event simulation or differential equations (including, but not limited, to finite-difference solutions). ``Sensors" or ``listeners" at each place can evaluate the chances and the number of times a given threshold of the number of tokens is crossed, or evaluate the time-averaged number of tokens at a given place. In the latter case the correlation matrix for all results can be evaluated as well, providing the mechanism for calculating the variances (in addition to the mean values) of global metrics that aggregate the readings of individual sensors.
 \item 
 Hierarchical constructions for combining multiple subnets are used to model large-scale systems. An a utomatic generation of  layers of multiple pages  that contain similar but possible distinct subnets is facilitated.
 \item In particular, an automatic ``color shift" is used to differentiate between the tokens from a given layer and ensure that the tokens enter the desired layer when needed.

\end{enumerate}

\subsection{Matching cars and customers example}
Figure~\ref{apn2a} A  depicts an APN model for the example discussed above.  An inhibitor originates at   the ``Driving" place, preventing two tokens from appearing in that place simultaneously (if one person drives, the other cannot drive the same car). An additional inhibitor that originates at the place ``Car broken" also terminates in that transition (if the car is broken, the customer cannot drive the car). Finally, if a customer drives a car (so there is a token in the place ``Driving"), and the car breaks (a token moves into the place ``Car broken"), then the enabler activates an immediate transition from ``Driving" to ``Waiting." In contrast, if the car is operational throughout the trip, the  transition from  the ``Driving" place to ``Not needed" is fired (the duration of the trip is specified as the delay of this transition).  

\begin{figure}[ht]
\begin{center} 
\includegraphics[width=5in]{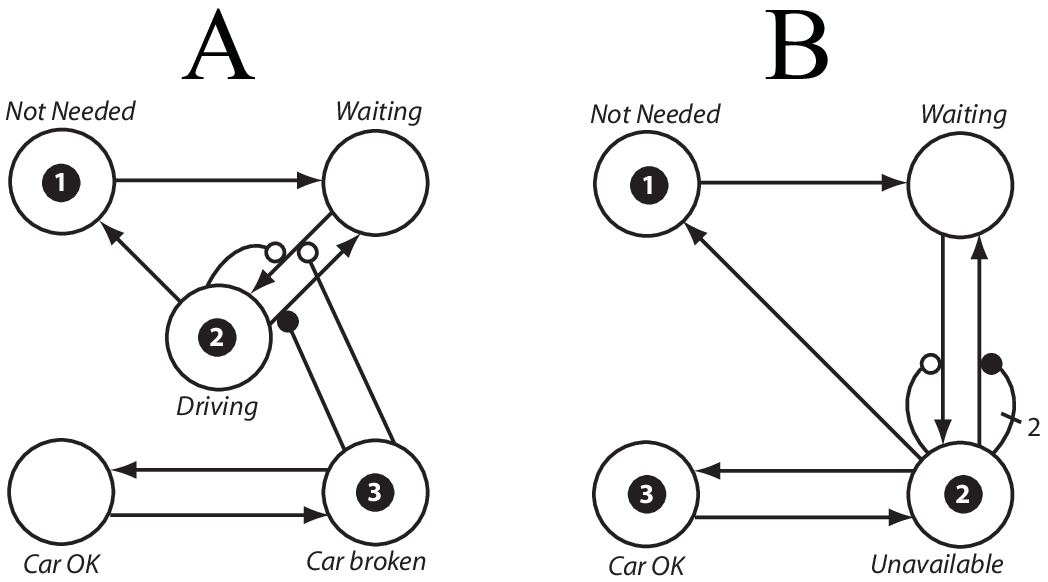}
 \end{center}
\caption{Using Abridged Petri Net diagram for two customers and one car: two alternative models }
\label{apn2a}
\end{figure}

An alternative model of the same process is shown in Figure~\ref{apn2a} B, demonstrating the Unavailablepotential tradeoffs in choosing the  modeling structure. The number of places is reduced to four by noting that  whether a  car is broken or  driven by another customer, the end result is that  that car is unavailable, and so it is possible to merge the corresponding two places into the ``Unavailable" place. Here color-specific policies of transitions ensure that the token representing the driver will return to the ``Not needed" place when the trip is completed, while the token representing the car will return to  the ``Car OK" place when the repair is completed. While those tokens are in the ``Unavailable" place, they all play the same role by displacing the other token that corresponds to driving.  While the number of places is reduced by one, transitions from the place ``Unavailable" have two color policies, so there seems to be no clear winner in terms of reducing the modeling complexity.

However, the latter model provides an interesting scaling capability. If there is more than one car, the model (after adjusting the appropriate trigger multiplicities and color policies) can also work correctly if customers have predetermined priorities, so that a customer with higher priority can  swap  a broken car for a working one, and a customer with the lowest priority will end up waiting for an available car if a car breaks (regardless of whether he or another customer drives that car).  The priorities are implemented by specifying small fixed-time delays for the transition from  the ``Unavailable" place  to the ``Waiting" place, ordered in such a way that customers with  lower priorities have smaller delays, and so the tokens of the corresponding color will be fired faster. 

For example, let us consider a situation where we have $m=3$ cars and $k=5$ customers, denoted with tokens of colors $1\ldots k$ ordered in such a way that the higher priority customers have higher-valued colors. The multiplicity of the inhibitor is $m$ (not allowing more than three tokens in the place) and the multiplicity of the enabler is $m+1$---initiating the removal of a token if it is ``extra."
Then by assigning small fixed delays for the transition from  the ``Unavailable" place  to the ``Waiting" place, so that for each $i$ the corresponding delay is $i \epsilon$, we ensure that correct priorities are observed (in an alternative implementation, those transitions can be explicitly specified as immediate with the possibility of specifying explicit priorities~\cite{Balbo2007}; the end result is effectively identical). Here the policies for the outgoing transitions from the ``Unavailable" place are assigned to make sure that tokens with colors $1\ldots k$ represent the customers (and so only the transitions to the ``Not Needed" and ``Waiting" are possible for those colors); and, similarly, colors $k+1\ldots k+m$ represent the cars (and  so the transition to the place ``Car OK" is ``blind" for tokens representing the customers). 

\begin{figure}[ht]
\begin{center} 
\includegraphics[width=2.5in]{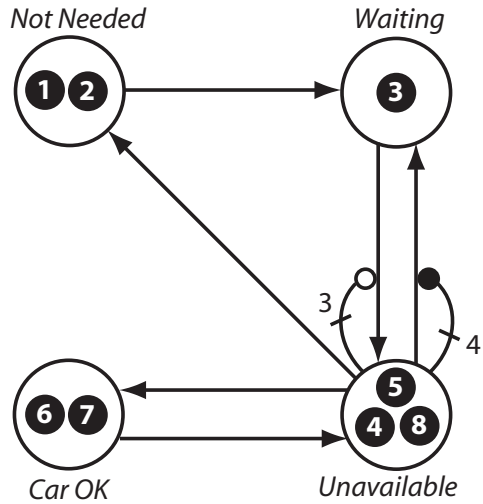}
 \end{center}
\caption{An Abridged Petri Net (APN) diagram for five customers and three cars}
\label{apn2c}
\end{figure}

While  this model is compact and relatively simple, it does not provide the resolution for tracking a specific pairing between the customers and the cars.\footnote{A customer who loses the utility of a car regardless of which car breaks is effectively specified in advance.} This resolution might be desirable if swapping of the cars in the middle of the trip is not allowed, and we want to track which customers' trips got interrupted.  In order to implement this aspect of behavior of the modeled system, we need to introduce a special place for each pairing. Since the number of such places should be at least equal to the number of cars, the resulting net can look quite complex, unless hierarchical models are implemented, as described in the next section.
\subsection{Periodic inspection example}
The following example is taken from a recent paper devoted to the issue of making Petri nets easier to understand in the context of system reliability modeling~\cite{Signoret2013}, where the example is used to demonstrate how a relatively simple system requires  fairly complex modeling using SPNs. As demonstrated below, the corresponding APN is significantly smaller and easier to read.

 There are two redundant components with failures only revealed by periodic tests; failures remain undetected until the next periodic test occurs.  The periodic test interval is $\tau$, and both components are tested at the same time. Once the failure of a component  is detected by a periodic test, the component is repaired, provided that a repair crew is available. There is only one repair crew, so both components cannot be repaired at the same time. If both components need repairs, one of them  is placed in a queue where it needs to wait until the repair of the other component is completed. In this example,  one of the components  is assigned a higher priority, so it is selected for repair first if both components need repair at the same time; however, there is no preemption of the repair process---if the repair of the lower priority component has started, it is completed regardless of the status of the other component. Alternative queueing policies can be easily implemented as well (see the next example).  

\begin{figure}[ht]
\begin{center} 
\includegraphics[width=4.0in]{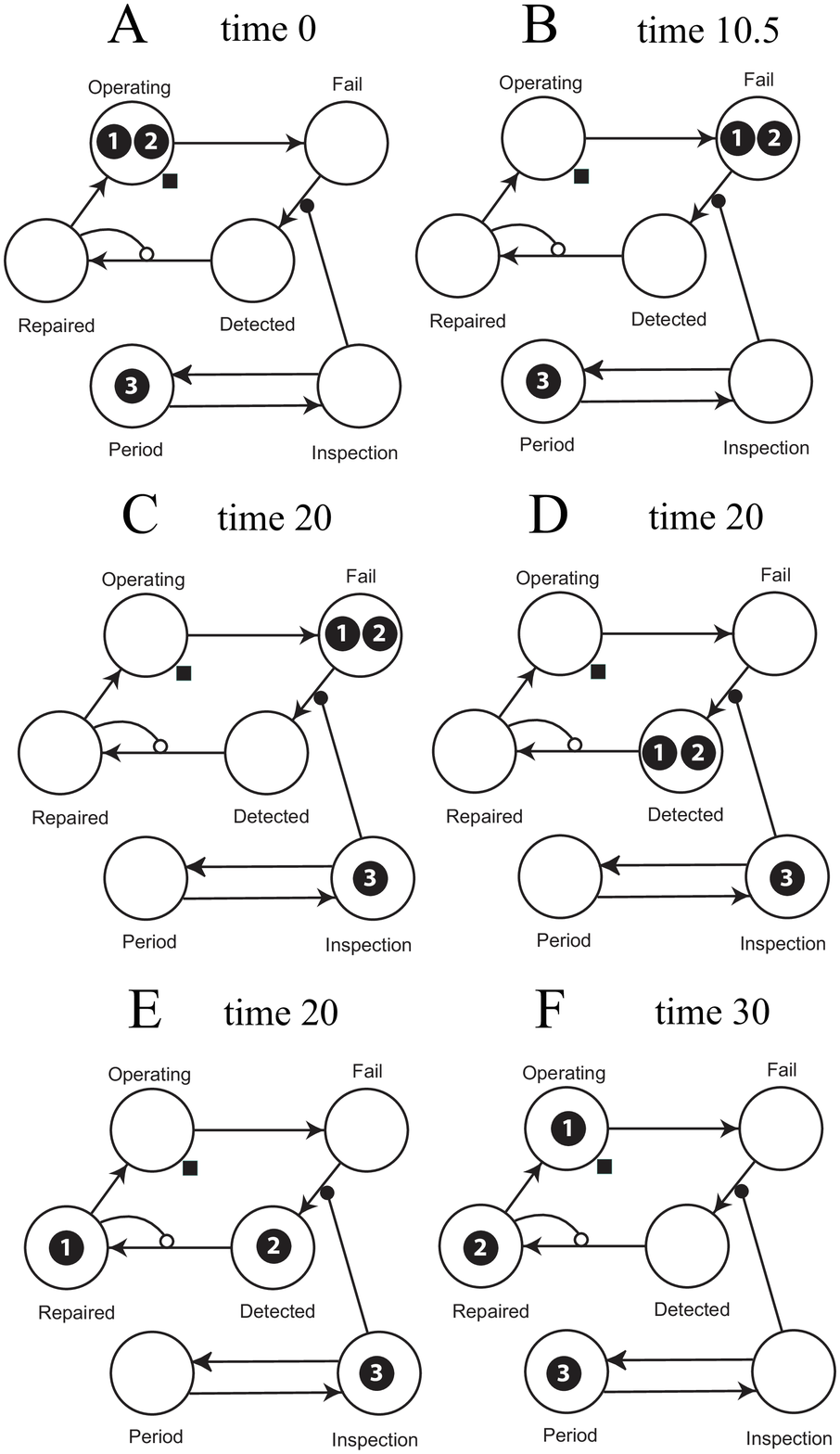}
 \end{center}
\caption{An Abridged Petri Net (APN) diagram two-component system with periodic inspection and limited repair resources }
\label{signoret1}
\end{figure}

 Figure~\ref{signoret1}  shows the  changes to the system as the simulation time progresses.  Frame $A$ in Figure ~\ref{signoret1} depicts the initial state of the model at time $t=0$.  There are two tokens in the  ``Operating" place. The transition from that place to the  ``Fail"  place is color-specific, so different failure distributions can be assigned to individual tokens that represent the components of the system. 
 In the considered example, the first component id denoted with Color $1$  with  transition properties following exponential distribution with the mean time to failure of  $100$  time units, while the failure of the  second component (Color $2$) follows Weibull distribution with  the scale parameter $\theta=200$ time units, and the shape parameter $\beta=2$. In a computer implementation of Monte Carlo simulation (see for example, \cite{Ross2012}), different runs are distinct, accounting for the random nature of the system's behavior, and the appropriate statistical measures of system performance are aggregated over many simulation runs. 

The particular run shown in Figure~\ref{signoret1}  demonstrates the model's behavior when both components fail within the same inspection interval (in this example, the inspection interval is chosen to be $20$ time units, which is implemented by assigning a fixed $20$ units delay to the  transition from the ``Period" place  to the  ``Inspection" place). The first failure occurs at Time $3.5$, and so the  Token $1$ moves to the ``Fail" place (not shown). The second failure occurs at Time $10.5$ (Frame $B$), so at this point both components have failed, and those failures have not been detected. At Time $20$, inspection occurs (the Token $3$ representing the inspection moves from the ``Period" place to the ``Inspection" place in (Frame $C$). This enables the instantaneous transition from  the  ``Fail" place to the ``Detected" place, and both tokens move from the former place  to the latter place (Frame $D$) (in order for this transition to fire, its delay should be smaller than  the delay for  the transition from the  ``Inspection" place to the  ``Period" place).  Next, the token representing the component with the higher repair priority (the  token $1$ in this case) immediately moves to the ``Repairing"  place (Frame $E$), thus preventing the other token moving to the same place due to the inhibitor originating at the ``Repairing" place. There are different methods for implementing priorities, the simplest being selecting a smaller fixed delay for the colors with higher priorities: for example, we can assign  to the transition from the ``Detected" place to the ``Repairing" place a fixed delay of $1\times 10^{-6}$ time units for the tokens with Color $2$, and  $2\times 10^{-6}$ for Color $1$ (assuming that the transitions from  ``Fail" to ``Detected" are the same for both colors). This ensures that the Color $1$ token is always selected first. After the first unit is repaired 
(in this case, a fixed duration of $10$ time units is used) at Time $30$  Token $1$ returns to the ``Operating" place. The inhibitor from the ``Repairing" place  becomes disabled, and so  Token $2$ can move to the ``Repairing" place (Frame $F$).

Using Monte-Carlo simulation, one can obtain the performance measures of interest. For example, based on one million runs of the system for $200$ time units, we can observe that  at least one unit is operating  $98.1\%$ of the time, while  both units will be operating $73.9\%$ percent of the time; the repair crew will be engaged $12.7\%$ of the time, with  the expected number of started repairs being $2.12$ for the simulated time; at least one unit has failure undetected $13.9\%$ of the time;  and a component is waiting for repair less than one percent of the time.  Here and below, the time percentage of a particular state is evaluated by taking an average over the second half of the system operation (between Times $100$ and $200$). If higher performance is desired, the effectiveness of various improvements can be evaluated: for example, adding a repair crew (which is implemented by changing the multiplicity of the inhibitor to $2$ (or, given the fact that we only have two components, simply removing the inhibitor altogether), will result only in marginal improvement, by raising the chances of both units operating to $74.6\%$ (this is expected given the low chances of a component waiting for repair). On the other hand, keeping one repair crew while making the inspection intervals twice as frequent would increase the chances of both units operating to $78.9\%$.  
\subsection{Queueing: First-In-First-Out (FIFO) policy }
 The ability to model queues is fundamentally important for a system's performance evaluation, to the point that special queueing constructs were suggested to enhance the  modeling capabilities of SPNs~\cite{Bause1993}. Here we show how one of the  most commonly used priority schemes, First-In-First-Out (FIFO), can be implemented using APNs (see Figure~\ref{fifo1}). At Time $t=0$ (Frame A), there are four tokens in the  ``Serviced" place representing four units, each with a distinct color number for visual identification. This is a closed queueing system, so serviced units return for service at some point in time (an open system can easily be implemented using tokens' sources and sinks). At some point tokens move to  the ``Needs Service" place (see Token $3$ in Frame C ), in which the tokens start aging in accordance with the properties of the transition to the ``Aging" place. This transition is chosen to be slow enough so that it never actually fires: instead, its sole purpose is to change the ages of tokens that are located in the ``Needs Service" place---the longer they stay in that place, the greater  the age. When the transition from the the ``Needs Service" place to the ``Servicing" place is enabled (Frame E), the token that arrived first, which is Token $3$ (Frame C), accumulates greater age, and is fired first (Frame F). 
 The tokens' ages are reset when they leave the ``Needs Service" place (so that the age can be used for other purposes in the model, including when the token returns to the place). It might be desirable  not to reset the age (or reset  it  only partially) however, as it will result in a policy that takes into account previous waiting times, and services the one with the maximum accumulated waiting time.  By reversing the aging process, the Last-In-First-Out (LIFO) ordering scheme can be implemented as well.  
\begin{figure}[ht]
\begin{center} 
\includegraphics[width=4.0in]{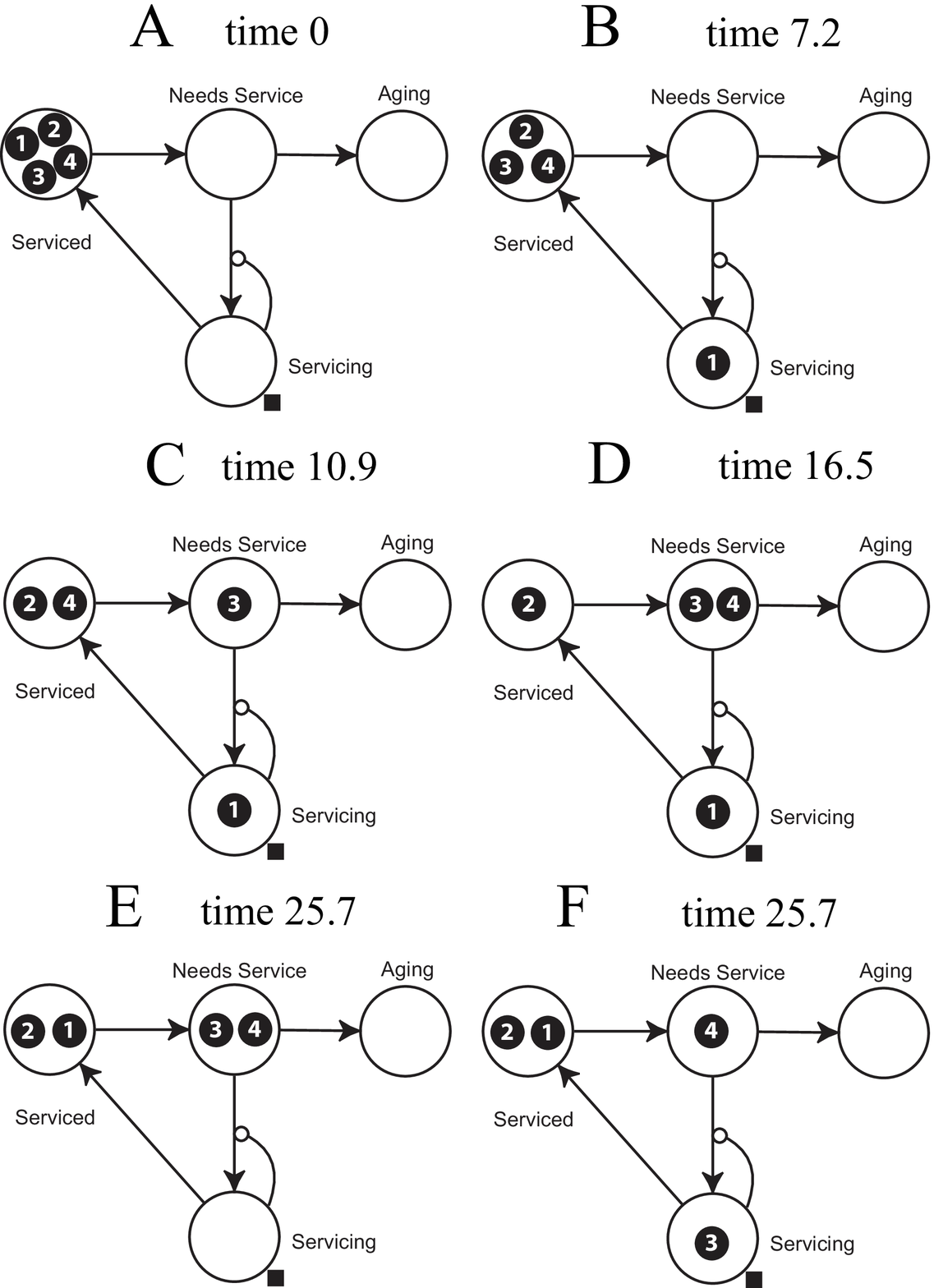}
 \end{center}
\caption{An Abridged Petri Net (APN) diagram of first-in-first-out (FIFO) queue}
\label{fifo1}
\end{figure}

\subsection{Air Traffic Control Example} 
 The next example pertains to modeling  air traffic control procedures~\cite{Calanni2013}.  Unlike the previous examples that were dealing with the interaction of supply and demand, here the main interaction of interest pertains to aircraft separation, which in APN notations translate into avoiding the situations where tokens representing aircraft occupy the same place simultaneously.
 A portion of the East Feeder Sector for LAX airport is modeled, where only the air traffic fluxes going through the feeder points ``GRAMM'' and ``LAADY'' are taken into consideration. Each airplane is assumed to fly an Optimized Profile Descent (OPD) that minimizes the use of the aircraft's throttle and follows a continuous descent profile instead of using a traditional stepwise descent. This approach  is attractive from the fuel-saving and noise-profile perspectives, but it  introduces uncertainty in terms of the  effective ground speed of the aircraft due to wind variability.  An underlying assumption for the feeder sector is that the feeder points and the merging point for the various traffic streams are positioned in such a way that airplanes flying from their respective feeder points will arrive at the merging point at the same time if moving with the same ground speed profile. Therefore, in nominal conditions, no conflict should arise if the aircraft are properly spaced within each traffic stream and the streams are properly synchronized, even though that is often not the case due to differences in the environmental conditions and the kinematics of each aircraft.

This synchronization between the streams is implemented by ATC that  checks for conflicts for every vehicle arriving at a feeder point; at that time, any aircraft trailing behind, either on the same traffic stream or on another one, is issued a $-20\,$kts or $-40\,$kts velocity-\-change if the horizontal distance from its feeder point is within $5\,$nm or $2.5\,$nm, respectively. 
This model captures only a portion of the actual procedures that ensure the proper spacing upon merging. A simple conflict resolution logic consisting of one single maneuver being issued per conflict was investigated, with only two maneuvers able to be executed. 

\begin{figure}[ht]
\begin{center} 
\includegraphics[width=4.0in]{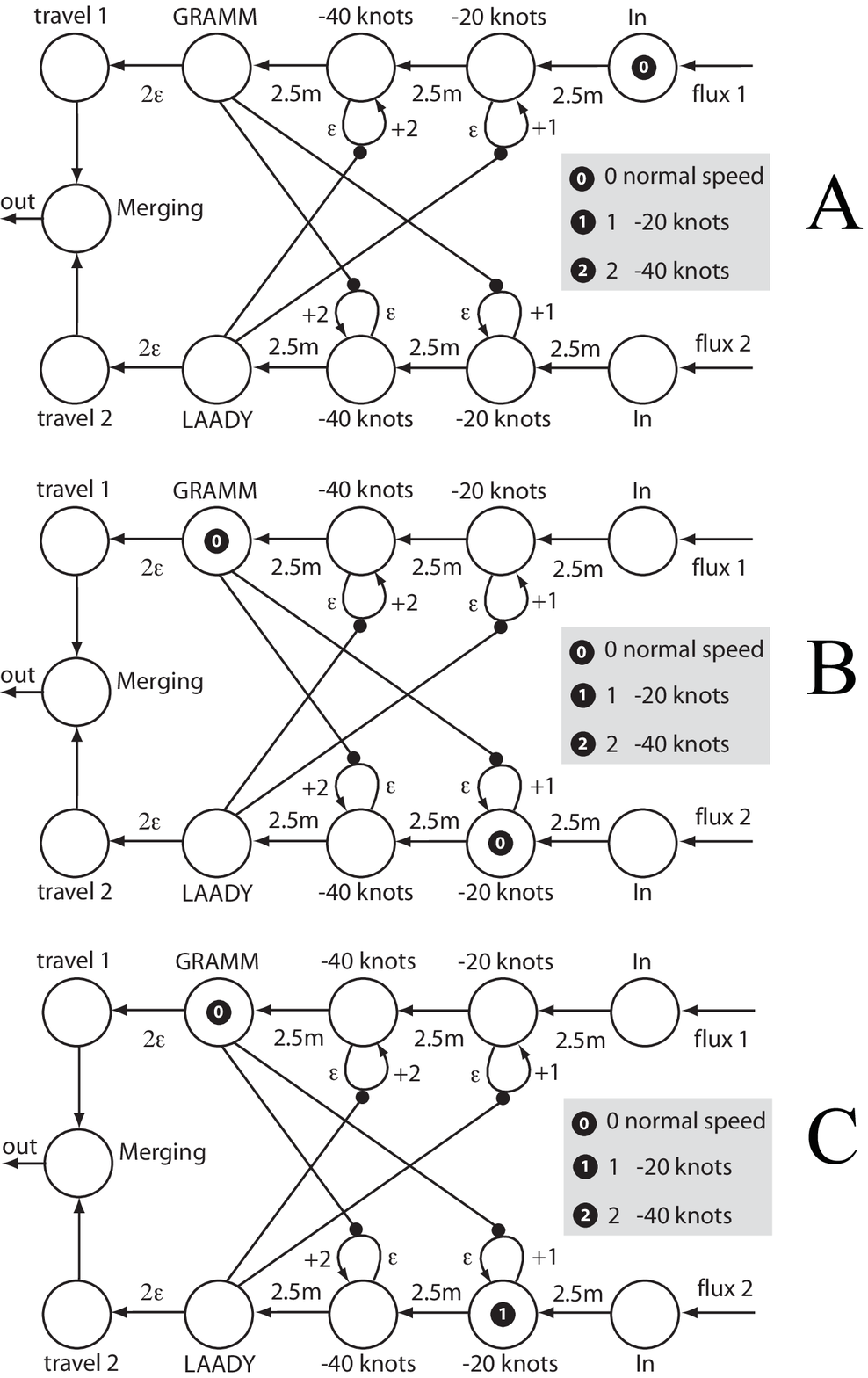}
 \end{center}
\caption{An Abridged Petri Net (APN) diagram of merging two aircraft streams east of Los Angeles (Frames A-C)  }
\label{merging}
\end{figure}

\begin{figure}[ht]
\begin{center} 
\includegraphics[width=4.0in]{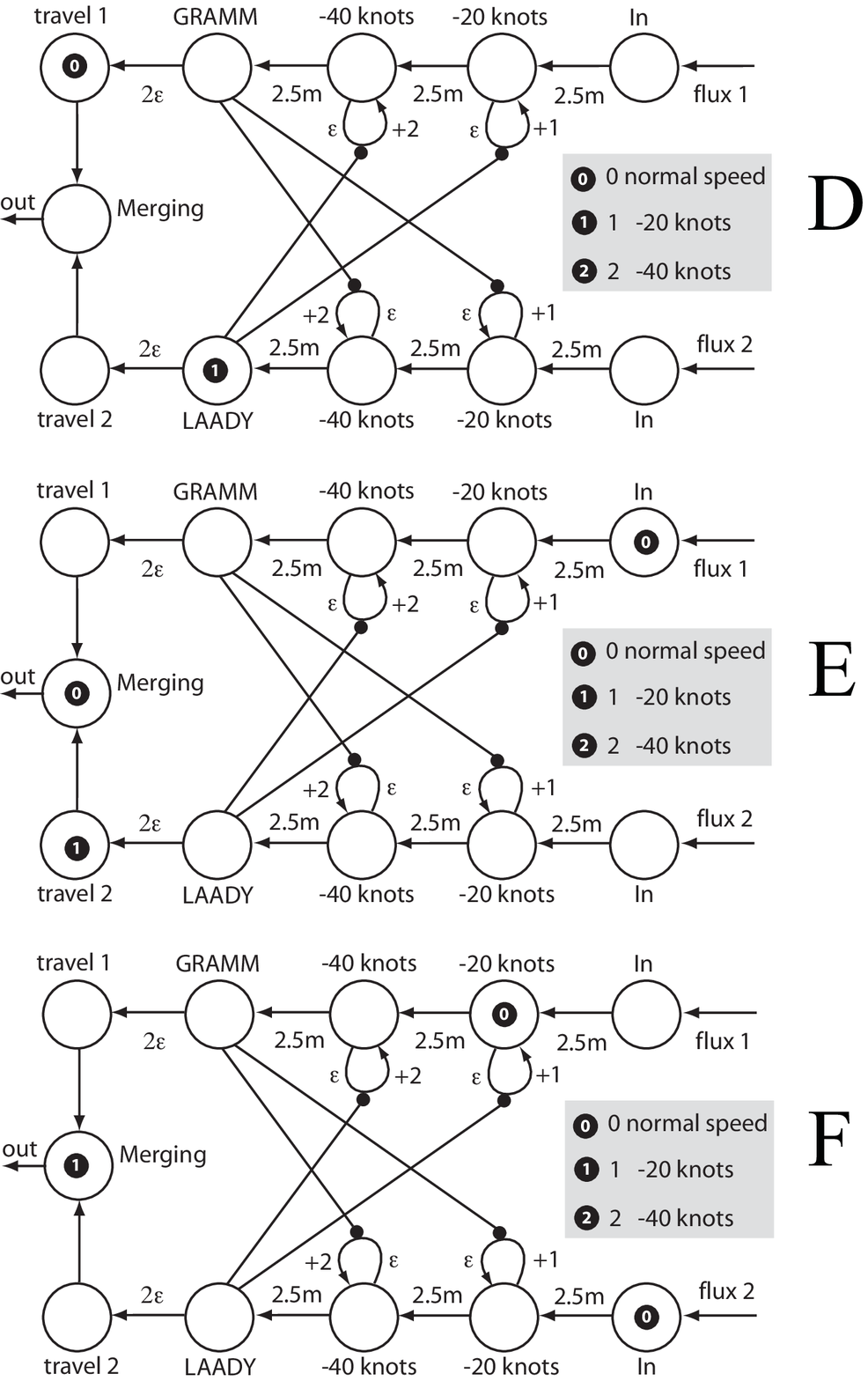}
 \end{center}
\caption{An Abridged Petri Net (APN) diagram of merging two aircraft streams east of Los Angeles (Frames D-F)  }
\label{merging2}
\end{figure}

The corresponding schematics of the APN model are  shown in Figures~\ref{merging} and~\ref{merging2}.  The APN model is obtained by the direct recasting  of the SPN model described in~\cite{Calanni2013}. Places  depict the possible positions of an aircraft, while the aircraft are represented by tokens. 
The top of the model represents  Flux 1 (moving from right to left)  passing through the GRAMM feeder point, while the bottom corresponds to LAADY Flux 2. Each token representing an aircraft is assigned one of  three colors (0 -- green, 1 -- yellow, 2 -- red) based on whether ATC issues a command for this aircraft to slow down. The delays associated with the transitions ``Travel 1" and ``Travel 2" are color-dependent, with the corresponding distributions provided by the agent-based simulation ~\cite{Calanni2013}. In Frame A (Figure~\ref{merging}), an aircraft appears in the  ``In" place of Flux 1. 

When a token is in  the GRAMM place (that is, an aircraft that has just reached the feeder point GRAMM,  as depicted in Figure \ref{merging}, Frame B), two transitions are enabled for the LAADY flux by means of enablers.  As a result, if there is a token in the input place for one of those transitions (Figure \ref{merging}, Frame B; this is true for the ``-20 knots"  place, {\em i.e.},  an aircraft is located somewhere between 2.5 miles and 5 miles away from the fix LAADY), then this token is fired through the transition ``+1," which deposits the token into the same place but changes its color from $0$ to $1$ (Figure \ref{merging}, Frame C).   Note that there is a fixed small delay $\epsilon$ associated with this transition while the token in  the GRAMM place stays for $2 \epsilon$ ensuring that this change occurs. The color-changing transition is  color-sensitive as well (the firing occurs only for the color $0$). Similarly (not depicted), if there is a token in the  ``-40 knots" place (e.g,  an aircraft  is located less than  2.5 miles from the fix LAADY), then this token is fired through the transition ``+2," which deposits the token into the same place while changing its color from $0$ to $2$. 

The reciprocal policies are implemented for the aircraft that need to be slowed down within the GRAMM flux based on their position when another aircraft passes the LAADY feeder point. The delay associated with the ``spacing" transition determines the degree of spacing violation: statistics are collected of the frequency of any two tokens located in the ``Merging"  place at the same time, so the longer delay of  the ``Spacing" transition  corresponds to the larger spacing. Effectively, here the spacing is evaluated in the time domain, which is obtained from the space domain based on the speed profile of the vehicles. In the considered run (Figure~\ref{merging2}), no spacing violation takes place, as the aircraft from Flux 1 appears in the ``Merging"  place prior to the aircraft from  Flux 2 (Frame E), and leaves that place before the latter aircraft gets there (Frame F). 

\subsection{Workflow example}
Next, let us consider an APN model for the processing of complaints~\cite{Aalst1998}.  Figure~\ref{complaints}  depicts the APN model that directly corresponds to the equivalent SPN model (~\cite{Aalst1998}, Figure 8.) Here the  transitions' names identify the  associated processes, and so the diagram is self-explanatory. It suffices to note that a token must be present in the $c_5$ place (indicating that either the questionnaire has been processed, or the time allocated for the questionnaire has elapsed)  in order for the ``processing complaint" and ``archive" transitions to be enabled and the corresponding activities to commence. The enabler to the ``archive" transition is not redundant in the case when no processing is needed (the transition from the $c_4$ place to the $c_6$ place fires).
\begin{figure}[ht]
\begin{center} 
\includegraphics[width=5.0in]{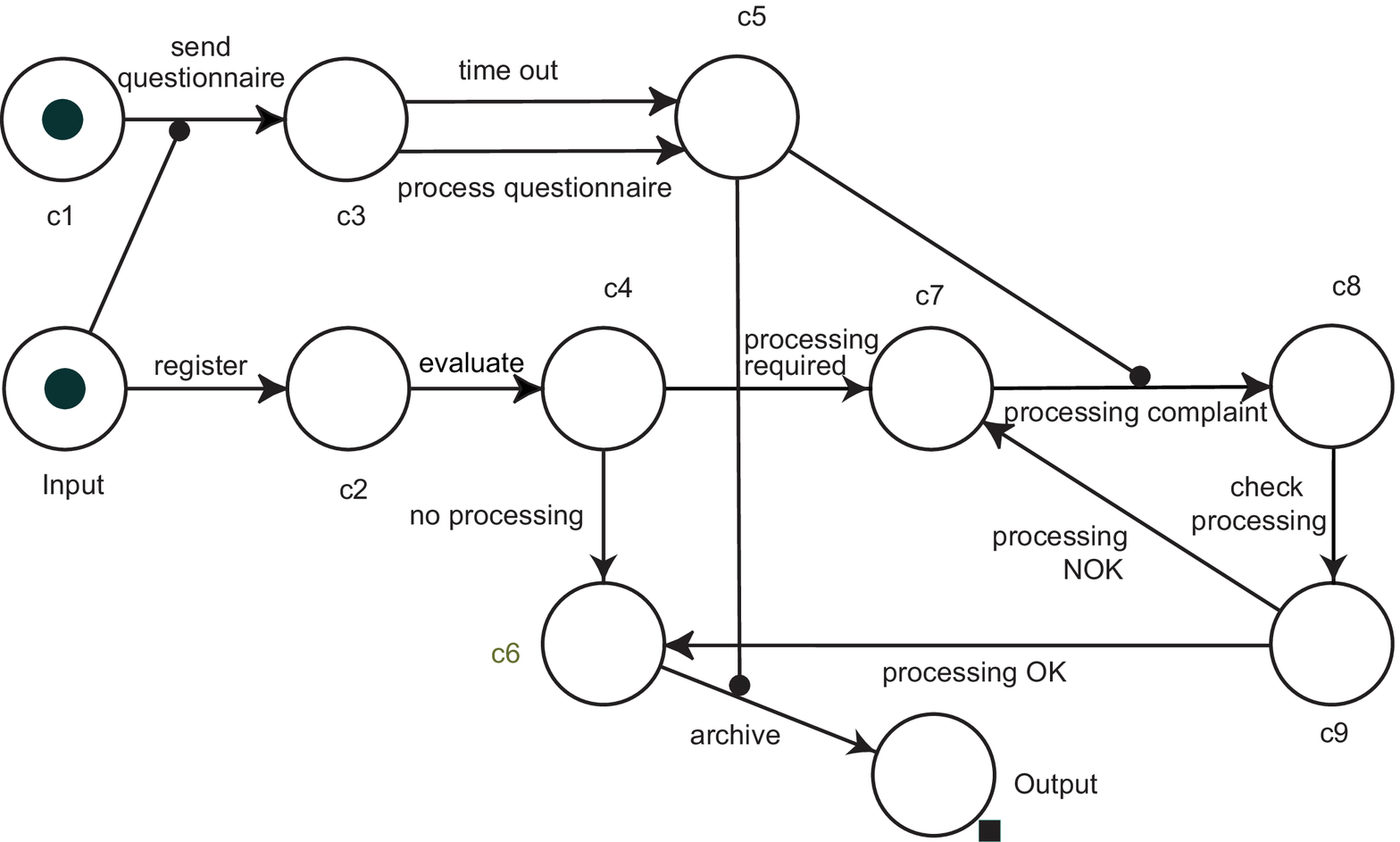}
 \end{center}
\caption{An Abridged Petri Net (APN) diagram of processing complaints  }
\label{complaints}
\end{figure}

\section{Hierarchical construction}
Let us return to the first example, and focus on matching pairs of supply and demand in a compact fashion. 
\subsection{Explicit matching of supply-demand pairs}
To this end,  it is convenient to expand the model into the third dimension (or ``depth") by considering subnets stacked in layers, as depicted  in Figures~\ref{apn3a}--\ref{apn3d}.  Unless we want to assign specific properties to particular pairs of tokens (see the next section), the number of layers does not need to exceed the number of demand-supply pairs  at any given time, so  that for $k$ cars and $n$ customers,  at most $\min\left\{k,n\right\}$ subnets are needed. 

In a software implementation, these layers can be created automatically, and one can switch the views among layers (by bringing the layer of interest to the top of the stack) as desired. The six triggers of each subnet (see the right, shaded part of Figure~\ref{apn3a}) are arranged to match one car with one customer within each net (assuming that both the customer and car are available). Frame A (Figure~\ref{apn3a})  depicts a situation where there is a customer waiting for a car, but no cars are available:  Token $1$ appears in `the `Waiting" place  and  enables an immediate firing of a token representing a car from the ``Available" place  to the ``Car used" place, but there are no tokens in the ``Available" place. At the time shown in Frame B, Ttoken $3$ representing a car moves the ``Available" place (the car has been repaired). Immediately, Token $3$ is moved to the ``Car Used" place (Frame C), and the inhibitor  from that place disables this transition, preventing more tokens (representing cars) from being fired (if there were more tokens in the ``Available" place, they would not move). It must be noted that the reverse transition is also enabled momentarily, so either that transition needs to be assigned  a lower priority (e.g., slower, if we are using fixed small delays), or we can make the triggers color-sensitive (so that the enabler in the transition from the ``Waiting" place to the ``Car Used" place  is sensitive only to car tokens, while the inhibitor is sensitive only  to the customer tokens---the multiplicity of the inhibitor should be equal to unity in this case). 

At the same time,  the transition for the customer to move from  the ``Waiting" place to ``Car Used" gets enabled and fires (token $2$, Frame D, Figure~\ref{apn3b}), and similarly the inhibitor ensures that only one token moves (since the total number of tokens in the ``Car Used"  place is two---that is, the multiplicity of the corresponding inhibitor). As a result, the customer represented by Token  $1$ continues to wait for a car (here no FIFO policies were implemented).
When the second car is repaired, transitions for Subnet $\#2$ are enabled (Frame E), and the matching of the customer with a car takes place (Frame F). At this point we have both customers driving cars.  If a car breaks in the first subnet (Frame G, Figure~\ref{apn3c}), the corresponding Token $3$ moves to the ``Car broken" place, then the transition from the``Car Used" place to the ``Waiting" place becomes enabled and  fires the token $2$ (Frame H).

When  the trip ends for the second pair (the customer's Token $1$ moves to the ``Not needed" place, Frame I ), then  the car's Token $4$ moves back to the ``Available" place (Frame J, Figure~\ref{apn3c}), enabling the possibility for the first customer to complete the trip (Frames K and L).

\begin{figure}[ht]
\begin{center} 
\includegraphics[width=4.5in]{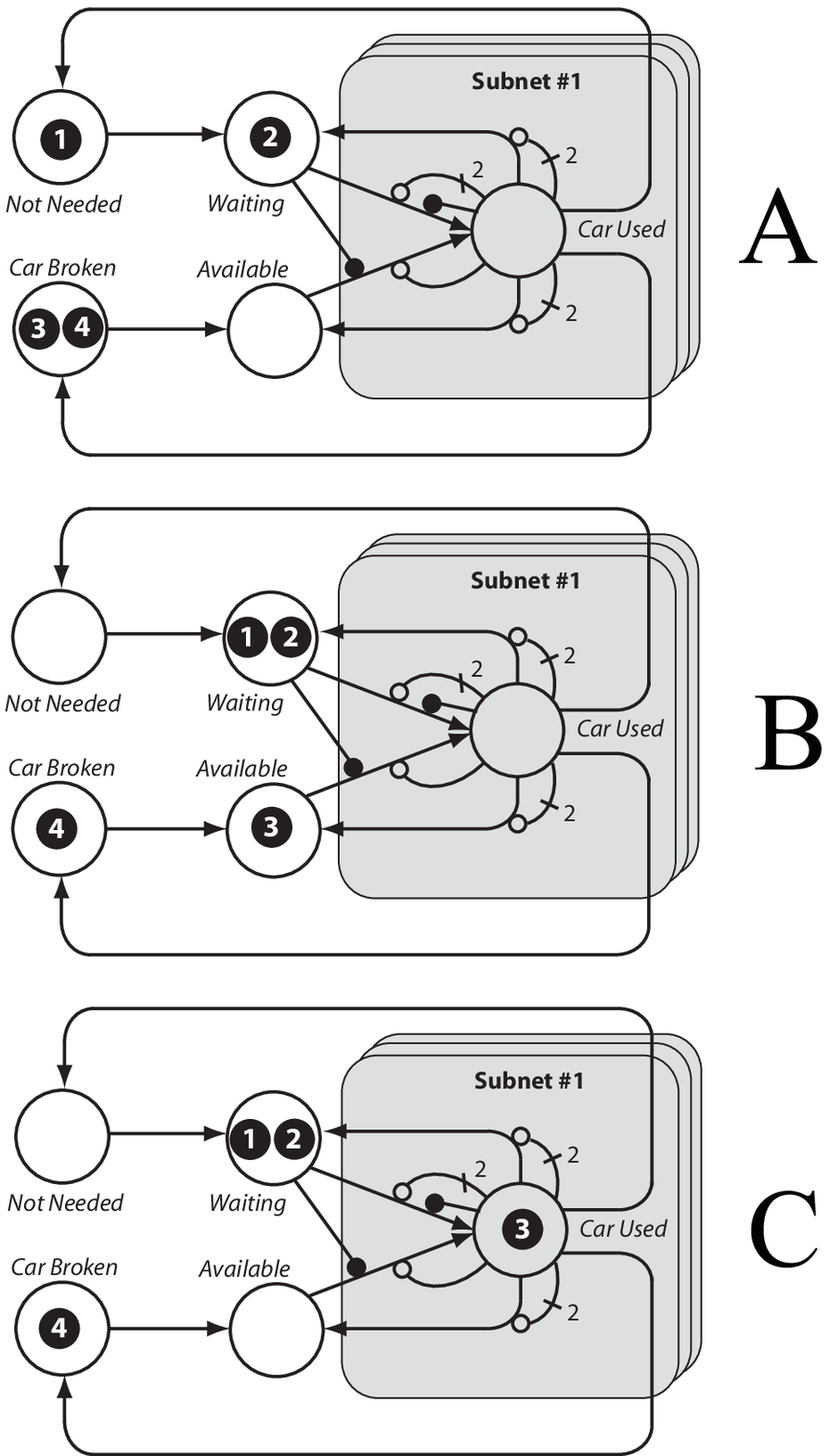}
 \end{center}
\caption{An Abridged Petri Net (APN) diagram for multiple cars and customers while tracking specific pairing of customers with the service (Frames A-C) }
\label{apn3a}
\end{figure}

\begin{figure}[ht]
\begin{center} 
\includegraphics[width=4.5in]{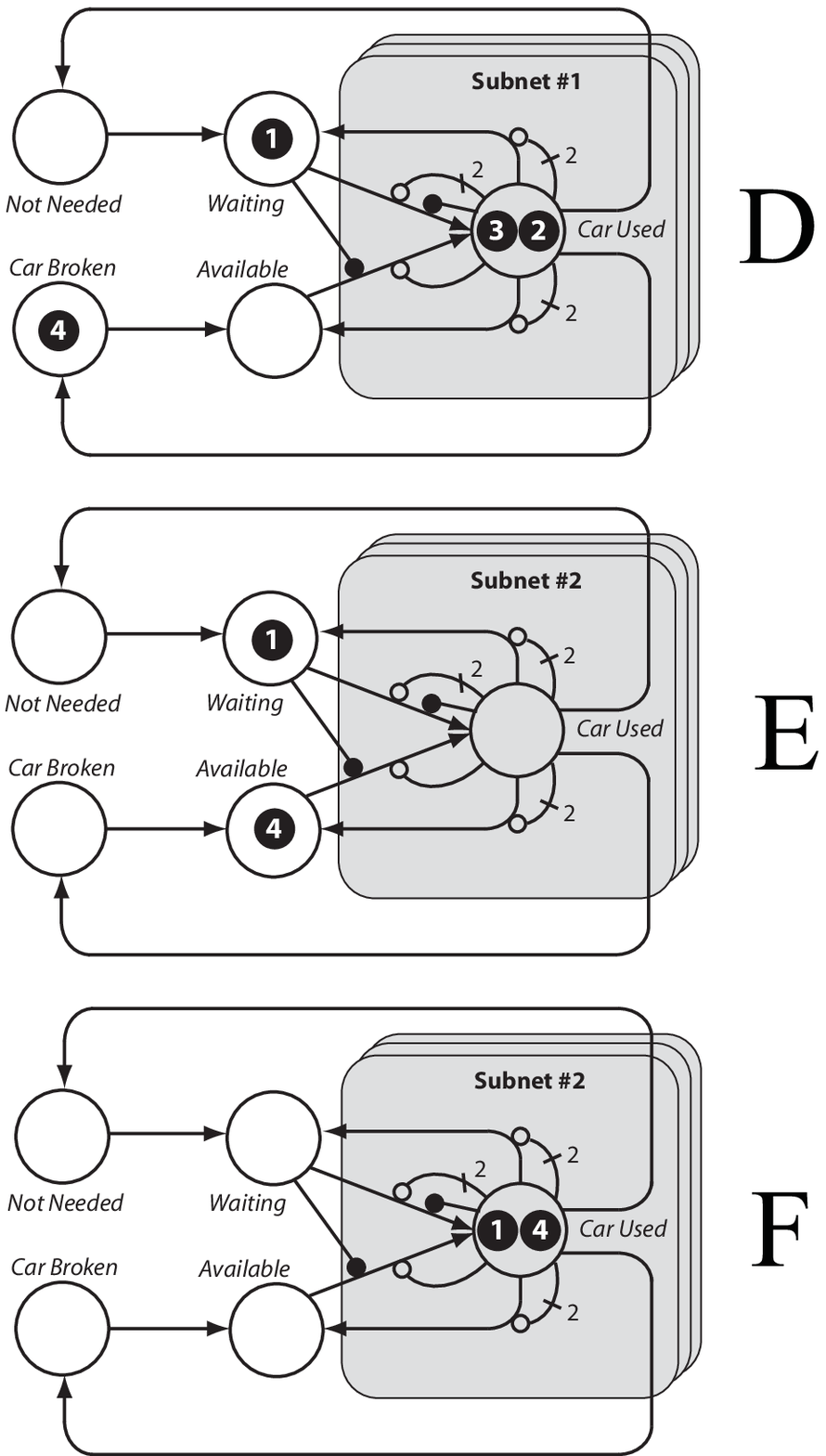}
 \end{center}
\caption{An Abridged Petri Net (APN) diagram for multiple cars and customers while tracking specific pairing of customers with the service (Frames D-F) }
\label{apn3b}
\end{figure}

\begin{figure}[ht]
\begin{center} 
\includegraphics[width=4.5in]{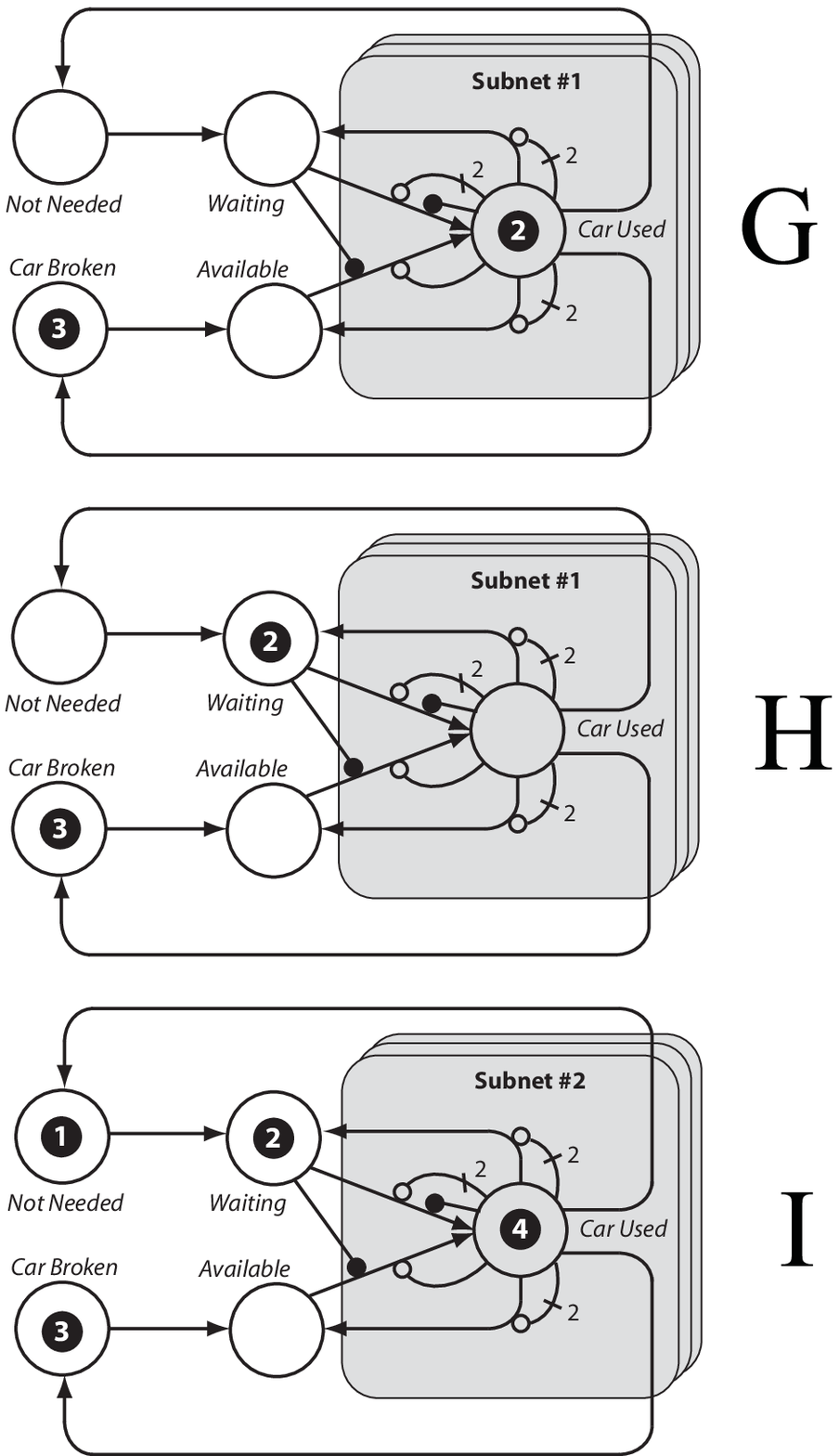}
 \end{center}
\caption{An Abridged Petri Net (APN) diagram for multiple cars and customers while tracking specific pairing of customers with the service (Frames G-I) }
\label{apn3c}
\end{figure}

\begin{figure}[ht]
\begin{center} 
\includegraphics[width=4.5in]{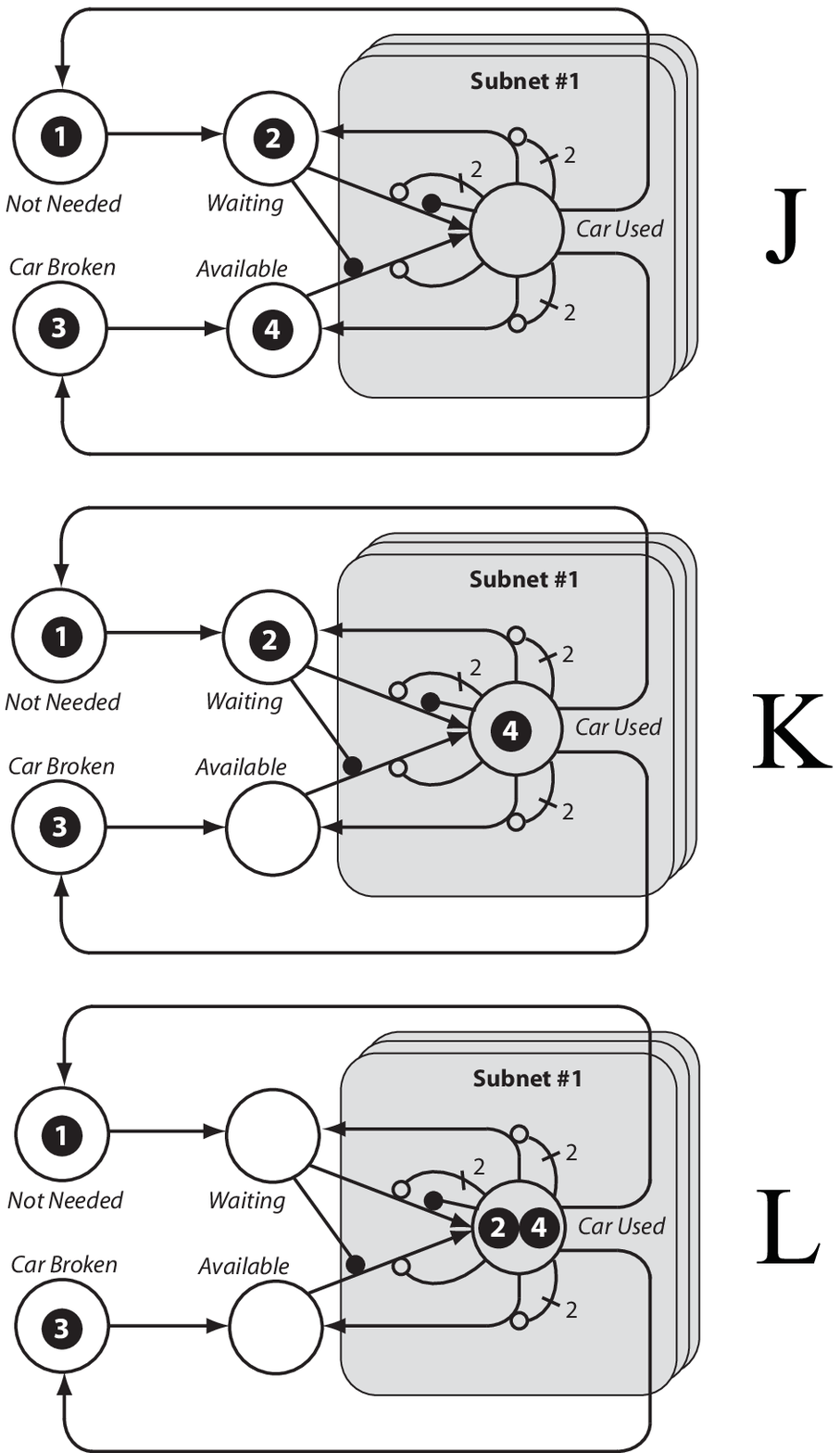}
 \end{center}
\caption{An Abridged Petri Net (APN) diagram for multiple cars and customers while tracking specific pairing of customers with the service (Frames J-L) }
\label{apn3d}
\end{figure}

\subsection{Automatic color shift}
In the previous example, all subnets were essentially equal---they simply provided a means of accounting for  each pair of tokens separately. However, one can further refine the model and allow for the possibility of differentiating among the subnets. In the considered example, we can assign a particular combination of colors to a specific subnet that might have distinct properties (for example, if a teenager drives a car, the chances of the car's breaking might increase).  
Next, we describe a general mechanism that facilitates the modeling of this and similar situations. 

Let us consider a situation with multiple but distinct subnets, so that tokens that leave a subnet should be returned to that specific subnet and not  to any other  subnet.  A repair process is one example of this situation, and so is the processing of documents: when a car is sent to the shop, one hopes to receive the same car fixed (and not somebody else's fixed car). Using an automated color shift allows this situation to be modeled automatically. First, we determine the range of colors utilized in the subnet that is used as a template for multiple subnets (see Figure~\ref{apn4}). Let us denote this range as $1\ldots j$ without the loss of generality (recall that colors are useful only to differentiate transition policies, so they can always be shifted together). 

If we want to create $m$ subnets, then we simply introduce a shifted range of colors: for $k$-th subnet ( $1\leq k \leq m$ ) the corresponding range of colors will be $jk+1\ldots j(k+1)$. An automatic check of color changes can be implemented to verify that colors are not changed outside of the subnet, so that a token cannot be accepted by the ``wrong" subnet. Frame A of Figure ~\ref{apn4} depicts subnet $\#1$ (color range $0\ldots2$) when the component  denoted with the token $2$ needs service. Frame B depicts Subnet $\#2$ (color range $3\ldots5$) when the component denoted with Token $3$ returns from service. Similarly, Frame C depicts the situation where Token $2$ returns to Subnet $\#1$, while Token $5$ from Subnet $\#2$ is in service.
\begin{figure}[ht]
\begin{center} 
\includegraphics[width=4.5in]{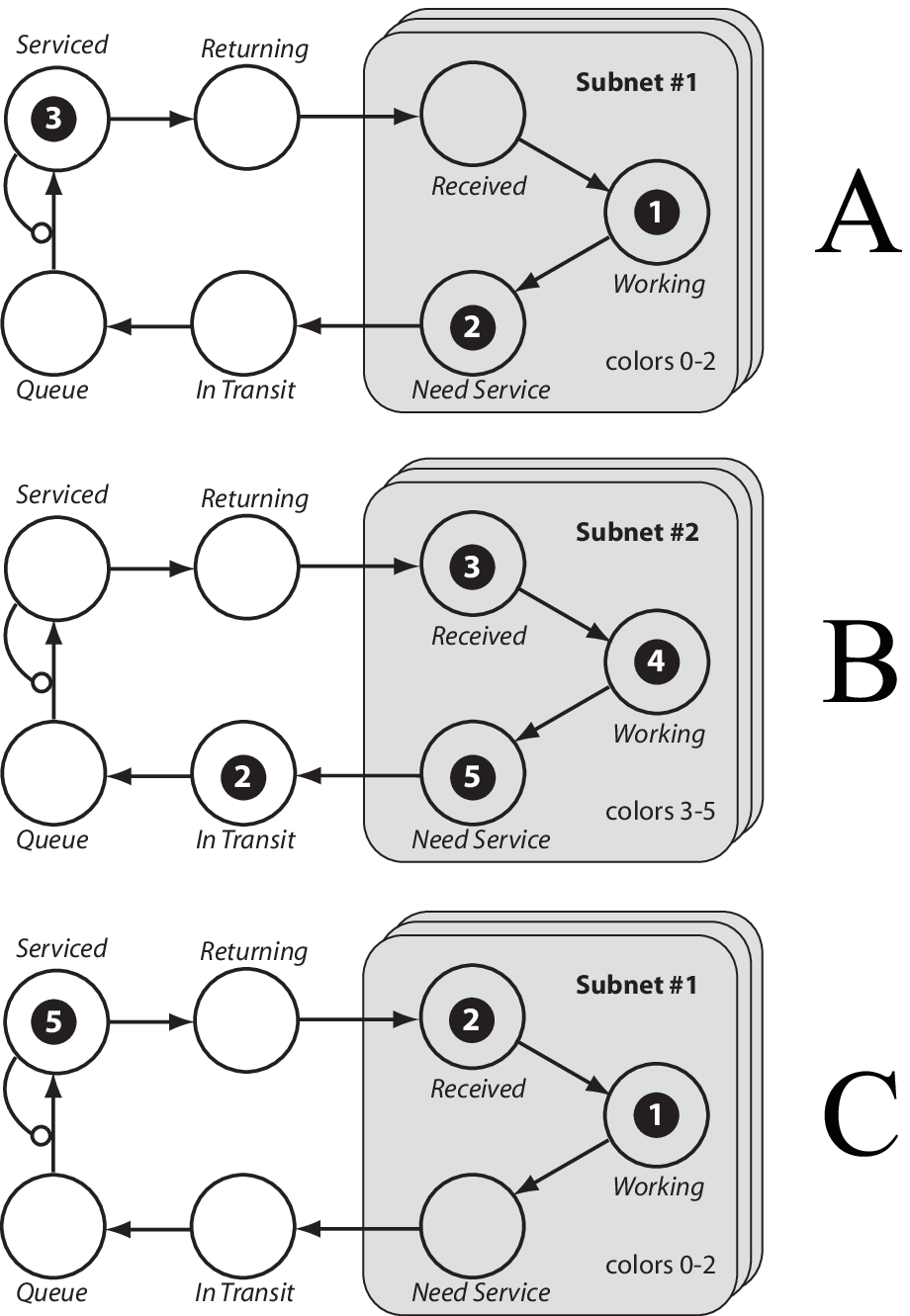}
 \end{center}
\caption{A general representation of the color shift model using Abridged Petri Nets (APNs). There are multiple transitions from the outside of the subnets to the subnets, but each transition is restricted in terms of the range of colors it can transmit to ensure that tokens return to their ``home" subnet }
\label{apn4}
\end{figure}

 \section{Conclusions}
A new graphical framework, Abridged Petri Nets (APNs) for modeling  the dynamics of complex stochastic systems, is presented.  APNs  derived from  Stochastic Petri Nets (SPNs) inasmuch as they both rely on component-based representation of system state space, in contrast to  Markov chains that explicitly model the states of an entire system. In both frameworks, the tokens  represent individual entities comprising the system; however,  SPN graphs contain two distinct types of nodes (places and transitions), with transitions serving the purpose of routing tokens among places, so that a pair of place nodes can be linked to each other only via a transient stop, a transition node. In contrast, APN graphs link place nodes directly by arcs (transitions), similar to state-space diagrams for Markov chains, and separate transition nodes are not needed.  
 Tokens in APN are distinct and have labels that can assume both discrete values (colors) and continuous values (ages),  and those values can change during simulation. Component interactions are modeled in APNs using triggers, which are either inhibitors  or the inhibitors' opposites, enablers. Hierarchical construction of APNs  rely on using stacks (layers) of submodels with automatically matching color policies.  As a result, there is no loss of modeling power as compared to SPNs and, as demonstrated by means of several examples, the resulting models are often more compact and transparent, therefore facilitating more efficient evaluation of the performance of complex systems.

\bibliographystyle{elsarticle-num}
\bibliography{simplepn}

\end{document}